\documentclass[11pt,twoside
]{article}

\usepackage{baaa2010}
\usepackage{graphicx}
\usepackage{rotating}
\usepackage{psfrag}
\usepackage{amssymb}
\usepackage[spanish,activeacute,english]{babel}
\usepackage[latin1]{inputenc}
\usepackage[T1]{fontenc} % Computer Modern (CM) fonts
\usepackage{ae,aecompl} % and: dvips -Pcmz -Pamz macros_aaa.dvi
\usepackage{latexsym}
\usepackage{verbatim}
\usepackage{amsmath}
\usepackage{amsfonts}
\usepackage{amssymb}
\usepackage{wasysym}
\usepackage[colorlinks=true,dvips]{hyperref}
\begin{document}
%%%%%%%%%%%%%%%%%%%%%%%%%%%%%%%%%%%%%%%%%%%%%%%%%%%%%%%%%%%%%%%%%%%%%%%%%%
%%%% SELECCIONE EL IDIOMA EN QUE SE ESCRIBE EL ARTÍCULO:              %%%%
%\myselectspanish
\myselectenglish
%%%%%%%%%%%%%%%%%%%%%%%%%%%%%%%%%%%%%%%%%%%%%%%%%%%%%%%%%%%%%%%%%%%%%%%%%%
\vskip 1.0cm
\markboth{Benaglia}%
{Energetics of stellar bow shocks}

\pagestyle{myheadings}
%%%% DESCOMENTE LA LINEA QUE DESCRIBE EL CARACTER DE SU TRABAJO       %%%%
\vspace*{0.5cm}
%\noindent TRABAJO INVITADO 
%\noindent PRESENTACIÓN ORAL
\noindent PRESENTACIÓN MURAL
%\noindent RESUMEN 
\vskip 0.3cm
\title{Energetics of nearby stellar bow shocks}

\author{Paula Benaglia$^{1,2}$}
\affil{%
  (1) Instituto Argentino de Radioastronom\'{\i}a (CCT-La Plata, CONICET)\\
  (2) Facultad de Ciencias Astron\'omicas y Geof\'{\i}sicas (FCAG, UNLP)\\
}

\begin{abstract} The latest survey of stellar bow shocks (Peri et al. 
2012) lists 28 candidates detected at IR wavelengths, associated with 
massive, early-type stars up to 3 kpc, along with the geometrical 
parameters of the structures found. I present here some considerations 
on the energetics involved, after the estimation of stellar wind power, 
infrared flux, stellar bolometric luminosity and radio flux limits for 
each source. The best candidates for relativistic particle acceleration 
are highlighted. \end{abstract}

\begin{resumen} Se consideran los candidatos a {\sl bowshocks} listados 
en el relevamiento E-BOSS.v1 (Peri et al. 2012). Tomando como base los 
datos all\'{\i} publicados, se calcula la luminosidad del viento, la 
luminosidad bolometrica de la estrella, la luminosidad correspondiente 
al flujo WISE medido, y la luminosidad a 1.4 GHz. Se discute la 
posibilidad de que alguno de los candidatos a bowshocks sea un sitio de 
aceleraci\'on de partículas a velocidades relativistas, con la 
consiguiente eventual emisión a altas energ\'{\i}as. \end{resumen}

\section{Introduction}

Runaway stars move with a speed larger than that of the average of the 
surrounding media, and tend to sweep the material found in the direction 
of motion. Ideally, the piled-up matter resembles a bow shock, larger 
for stars with strong winds. The first systematic search for bow shocks 
was carried out using the 1988 IRAS database (see Noriega-Crespo et al. 
1997 and references therein). The authors listed $\sim$ 60 nebulosity 
sources around early-type stars, some arcmins in size. The advenement of 
new IR missions like Spitzer or WISE (2011) produced images at the 
arcsec resolution, and allowed not only to check former results but to 
address a much deeper search for stellar bow shocks.

With the help of the Tetzlaff et al. (2010) catalogue of runaway stars 
up to 3 kpc, we carried out a study to look for signatures of WISE 
emission (Wright et al. 2010) towards all nearby O-B2 stars. The results 
were compiled in the Extensive stellar BOw Shock Survey, version 1 
(E-BOSS.v1, Peri et al. 2012), that lists and describes $\sim$ 30 bow 
shocks.

Very recently, Benaglia et al. (2010) analyzed the possibility that 
stellar bow shocks can give rise to high-energy emission, by studying 
the surroundings of the O supergiant BD+43$^\circ$3654. We thus 
seeked for low-frequency radio emission from E-BOSS.v1 candidates, that 
could be indicative of synchrotron radiation from the stellar bowshocks, 
with the New VLA Sky Survey (Condon et al. 1998, 1.4 GHz). We found 
three new radio sources possible associated with E-BOSS candidates (Peri 
et al. 2012). Some comments on how common and under which conditions 
stellar bow shocks could be high energy emitters are presented here.

%----------------------------------------------------------------------------------

\section{Bow shocks as acceleration sites}

The bow shock of BD+43$^\circ$3654 is the prototype of the non-thermal 
runaway stars. It has been observed with the VLA at two frequencies (L 
and C band), and the spectral index distribution showed average values 
$\alpha \sim -0.4$ ($S \propto \nu^\alpha$), characteristic of 
synchrotron radiation. Benaglia et al. (2010) proposed for the first 
time that a stellar bow shock can host relativistic particles also 
involved in high-energy emission processes, and built a zero-order SED 
that fits the radio emission and predicts the detectability at shorter 
wavelengths.

The powerful stellar winds of the early-type stars interact with the 
ambient medium creating two shocks, separated by a discontinuity: a 
forward shock with a velocity near the stellar one, and a reverse one 
fast as the stellar wind (see Benaglia et al., these Proceedings).

Del Valle and Romero (2012) improved the basic model, and applied it to 
the case of $\zeta$ Oph, concluding that the gamma-ray emission would be 
weak --if compared with other non-thermal emitters- but still detectable 
by forthcoming instruments like the Cherenkov Telescope Array.

The luminosities involved in the phenomenon can be estimated if the main 
variables of the star+bow shock systems are known. The E-BOSS.v1 
database comprises stellar and bow shock parameters, like the stellar 
distance $d$, the wind terminal velocity $v_{\rm wind}$, the stellar 
mass loss rate $\dot{M}$, the tangential and radial stellar velocities,
%the lenght and width of the bowshock feature, 
the stand-off distance $R_0$ and the original ambient density $n_{\rm ISM}$ 
at the position of the star, of each candidate.

%The Mach number can be estimated as $M \sim \sqrt{\Delta / R_0}$, where 
%$\Delta$ is the width of the shocked stellar wind (del Valle \& Romero 
%2012). %Let $\Delta \sim w$; then the derived values of the Mach number 
%for each candidate %are given in Table 1.

\section{Luminosities}

Table 1 lists information on the stellar bow-shock candidates 
compiled in E-BOSS.v1, related with 28 OB stars. References on spectral 
types, distances -including error bars- 
and wind terminal velocities are given in Peri et al. 
(2012). Stellar velocities were computed from radial and tangential 
velocities.  Distance values derived from parallax measurements, and stellar 
velocities computed from radial velocity information and proper motions
are flagged. The large uncertainties in some stellar distances introduce 
important errors in $v_*$, $R_0$ and $n_{\rm ISM}$. The stellar luminosities 
are from Martins et al. (2005) and Benaglia et al. (2007).

% for those stars where the tangential velocities
%were computed from proper motions (see Tabel 1). 
%Distance values tagged with a ``:'' (colon) represent errors larger than 50\%.

The luminosity of the stellar wind is $L_{\rm wind} = 0.5 \dot{M} 
(v_{\rm wind})²$, and represents the available kinetic power. To ensure 
that the flow is compressible and shocks can develop, the magnetic 
energy density must be in subequipartition with respect to the kinetic 
energy $L_{\rm wind}$. In this case, the value of the magnetic field 
intensity in the flow can be expressed as $B^2 / 8 \pi = L_{\rm wind} / 
(v_{\rm wind} 4 \pi R_0)$ (del Valle \& Romero 2012). The authors 
considered that only a small fraction of the kinetic power available 
goes into relativistic particles $L_{\rm rel}$, and adopted a 10\% 
factor.

The infrared luminosity was estimated by measuring the WISE flux 
averaged over the extension of the bow shock, subtracting the background 
contribution, and applying the conversion factors (see 
wise2.ipac.caltech.edu).
%/release/prelim/expsup/sec2\_3f.html).
 
The bow-shock candidates that correlate with NVSS emission are HIP 11891, 
HIP 38430, HIP 88652 and BD+43$^\circ$3654. Their radio luminosity was 
derived by measuring the corresponding flux. For the rest of the 
candidates, an upper limit of 3$\sigma$ = 3 rms for the NVSS radio flux density 
was assumed.
% (see Table 1).

Both $L_{\rm WISE}$ and $L_{\rm NVSS}$ will also be affected by large distance
uncertainties. Distance errors of 50\%, for instance, corresponds to a 
factor 5 in luminosities.
%$L$.

\begin{sidewaystable}
\centering
  \begin{tabular}{l l l r r r r r r r r}
Star & Sp.type & $d$    & $n_{\rm ISM}$ & $R_0$ & $v_*$ & $v_{\rm wind}$ & $L_*$  & $L_w$      & $L_{\rm WISE}$ & $L_{\rm NVSS}$ \\
     &         &  (kpc) &(cm$^{-3}$)  & (pc)  & (km/s) & (km/s) &    log    & (erg & /    &  s) \\
      \hline
HIP 2036  & O9.5III+...    & 0.76$\pm0.16$ &	130	&0.22	&16.0	& 1200 & 38.7 &	35.3 & 36.6 & $<$27.5 \\
HIP 2599  & B1 Iae         & 1.50$\pm0.30$ &	0.4	&1.27	&26.3	& 1105 & 39.0 &	34.7 & 37.5 & $<$28.0 \\
HIP 11891$^\dag$&O5 V((f)) & 0.40$\pm0.15$ (1) & 17	&0.12	&49.5	& 2810 & 38.1 &	36.4 & 34.9 &    30.3 \\
HIP 16518 & B1 V           & 0.65$\pm0.16$ (1) & 0.2	&0.13	&53.5	&  500 & 39.1 & 32.7 & 37.2 & $<$27.3 \\
HIP 17358 & B5 III         & 0.15$\pm0.10$ (1) & 600	&0.04 	&(2) 35.2	&  500 & 38.0 &	31.9 & 35.5 & $<$26.1 \\
HIP 22783 & O9.5 Ia        & 1.60$\pm0.30$ &	0.02	&4.67	&(2) 52.4	& 1590 & 39.2 & 35.3 & 37.1 & $<$28.1 \\
HIP 24575 & O9.5 V         & 0.55$\pm0.07$ &	3	&0.06	&152.0	& 1200 & 38.3 & 34.7 & 35.0 & $<$27.2 \\
HIP 25923 & B0 V           & 0.90$\pm0.20$ (1)&	1	&0.39	&24.2	& 1000 & 38.1 & 34.3 & 35.8 & $<$27.6 \\
HIP 26397 & B0.5 V         & 0.35$\pm0.15$ (1)&	2	&0.10	&22.4	&  750 & 38.1 & 33.4 & 34.6 & $<$26.8 \\
HIP 28881 & O8 Vn          & 1.5 :         &	0.3	&1.85	&(2) 17.7	& 2070 & 38.5 & 34.6 & 36.2 & $<$28.1 \\
HIP 29276 & B1/2 III       & 0.40$\pm0.03$ (1)&	0.003	&0.23	&32.0	&  600 & 37.8 & 32.1 & 35.6 & $<$26.9 \\
HIP 31766 & O9.7 Ib        & 1.40$\pm0.03$ &	0.03	&0.82	&58.8	& 1590 & 39.1 & 35.9 & 37.2 & $<$28.0 \\
HIP 32067 & O5.5V((f))+... & 2.10$\pm0.40$ &	0.1	&1.85	&38.8	& 2960 & 39.0 & 35.6 & 37.3 & $<$28.4 \\
HIP 34536 & O6.5V((f))+... & 1.30$\pm0.20$ &	0.01	&1.5	&59.7	& 2456 & 38.8 & 35.6 & 37.7 & $<$27.9 \\
HIP 38430$^\dag$&O6Vn+...   & 0.9 : (1)     &	60	&0.13	&(2) 30.9	& 2570 & 38.9 & 36.2 & 35.7 &    31.3 \\
HIP 62322 & B2.5 V         & 0.11$\pm0.004$ (1)&	0.04	&0.03	&42.2	&  300 & 37.9 & 32.2 & 33.8 & $<$25.4 \\
HIP 72510 & O6.5III(n)(f)  & 0.35$\pm0.18$ (1)&	0.2	&0.15	&74.4	& 2545 & 39.1 & 35.7 & 34.2 & $<$26.8 \\
HIP 75095 & B1Iab/Ib       & 0.80$\pm0.50$ (1)&	40	&0.12	&28.9	& 1065 & 39.0 & 34.7 & 35.8 & $<$27.5 \\
HIP 77391 & O9 I           & 0.8 : (1)        &	30	&0.23	&(2) 24.2	& 1990 & 39.2 & 35.5 & 37.3 & $<$27.5 \\
HIP 78401 & B0.2 IVe       & 0.22$\pm0.02$ &	2	&0.39	&(2) 38.6	& 1100 & 37.8 &	34.7 & 36.8 & $<$26.4 \\
HIP 81377 & O9.5 Vnn       & 0.22$\pm0.02$ &	1	&0.32	&28.6	& 1500 & 38.3 &	34.2 & 35.3 & $<$26.4 \\
HIP 82171 & B0.5 Ia        & 0.85$\pm0.12$ &	1	&0.17	&84.6	& 1345 & 39.1 & 34.7 & 37.1 & $<$27.6 \\
HIP 88652$^\dag$&B0 Ia     & 0.65$\pm0.30$ (1)&	2	&0.28	&31.1	& 1535 & 39.0 & 35.6 & 35.9 &    30.1 \\
HIP 92865 & O8 Vnn         & 0.35$\pm0.12$ (1)&	0.003	& 0.31	&(2) 41.2	& 1755 & 38.5 & 33.6 & 34.6 & $<$26.3 \\
HIP 97796 & O7.5 Iabf      & 2.2 :         &	0.02	& 3.84	&(2) 110.4	& 1980 & 39.3 & 35.8 & 38.8 & $<$26.8 \\
HIP 101186& O9.7 Ia        & 1.50$\pm0.40$ &	0.1	&1.73	&35.8	& 1735 & 39.1 & 35.3 & 38.5 & $<$28.4 \\
BD+43 3654& O4 If          & 1.45$\pm$ 0.05&	0.2	& 1.48	&(2) 67.7	& 2325 & 39.5 &	37.1 & 35.7 &    31.2 \\
HIP 114990& B0 II          & 1.4 :         &	0.05	& 0.61	&(2) 135.7	& 1400 & 39.0 &	35.6 & 35.2 & $<$28.0 \\
\hline
\multicolumn{11}{l}{$\dag$: Potential gamma-ray emitters. They were chosen by IR luminosity, wind terminal velocity and NVSS flux.}\\
\multicolumn{11}{l}{(1): Stellar distances derived from parallaxes. (2): Stellar velocities derived from radial velocities and proper motions.}\\
\multicolumn{11}{l}{':': errors larger than 50\%.}\\
  \end{tabular}
  \caption{Stellar, wind and bow-shock luminosities of the 28 bow-shock candidates of Peri et al. 2012.}\label{tab:permiss}
\end{sidewaystable}
%}

\section{Discussion}  

The results presented in Table 1 allow to draw some conclusions in 
regards with the emission mechanisms acting at the different bow shocks. 
The Hipparcos stars \# 24575, 97796 and 114990 stand out with larger 
stellar velocities. The first one was studied at X-rays by L\'opez 
Santiago et al. (2012), who proposed the stellar bow shock as the first 
X-ray non-thermal emitter.

The faster stellar winds will be more efficient accelerators of 
relativistic particles, since the acceleration efficiency $\eta \propto 
(v_{\rm shock}/c)^2 \sim (v_{\rm wind}/c)^2 \leq 10^{-4}$.

IC scattering will be favoured wherever the infrared photon 
field is strong, i.e., for stars with larger $L_{\rm WISE}$. In the 
stellar bow-shock scenario, the stellar UV photon field is not relevant 
to IC, due of the large separation from the star.

According to Del Valle \& Romero (2012), electron synchrotron losses 
dominate above relativistic Bremsstrahlung. Convection governs proton 
losses.

A crucial factor for high-energy emission detection is the distance to 
the star+bow shock system, as the luminosity decays with $d²$.

Finally, the presence of significant radio emission at low frequencies 
(like 1.4 Ghz) is a very strong hint to look for relativistic particles. 
For those objects, observing campaigns at two or more radio frequencies 
should be implemented, in full polarization mode if possible, to study 
the radiation regime. If evidence of synchrotron emission is found, the 
conditions mentioned above can help to disentangle the importance of the 
different contributions to high energy emission.

\acknowledgements P.B. is always grateful to G.E. Romero for useful 
discussions, and thanks C.S. Peri for comments, and PICT 
00848-2007, ANPCyT.


\begin{references}

Benaglia, P., Vink, J. S., Mart\'{\i}, et al. 2007, A\&A, 467, 1265

Benaglia, P., Romero, G. E., Mart\'{\i}, et al. 2010, A\&A Lett, 517, 10

Condon, J. J., Cotton, W. D., Greisen, E. W., et al. 1998, AJ, 115, 1693

Del Valle, M. V. \& Romero, G. E. 2012, A\&A, 543, 56

L\'opez-Santiago, J. Miceli, M., del Valle, M. V., et al. 2012, ApJ Lett, 757, 6

Noriega-Crespo, A., van Buren, D., Dgani, R. 1997, AJ, 113, 780

Peri, C. S., Benaglia, P., Brookes, D., et al. 2012, A\&A, 538, 108

Tetzlaff, N., Neuhausser, R., Hohle, M. M. 2010, MNRAS, 410, 190

Wright, E. L., Eisenhardt, P. R. M., Mainzer, A. K., et al. 2010, AJ, 140, 1868

\end{references}
\end{document}